\documentclass[conference]{IEEEtran}
\IEEEoverridecommandlockouts
\usepackage{cite}
\usepackage{amsmath,amssymb,amsfonts}
\usepackage{algorithmic}
\usepackage{graphicx}
\usepackage{textcomp}
\usepackage{xcolor}

\usepackage[normalem]{ulem}

\def\BibTeX{{\rm B\kern-.05em{\sc i\kern-.025em b}\kern-.08em
    T\kern-.1667em\lower.7ex\hbox{E}\kern-.125emX}}
\begin{document}

\title{CAVEMOVE: An Acoustic Database for the Study of Voice-enabled Technologies inside Moving Vehicles\\
\thanks{The work presented in the paper was funded by an internal grant of the Institute of Computer Science of the Foundation for Research and Technology-Hellas (FORTH).}
}

\author{\IEEEauthorblockN{Nikolaos Stefanakis\textsuperscript{* \dag}, Marinos Kalaitzakis\textsuperscript{*}, Andreas Symiakakis\textsuperscript{*}, Stefanos Papadakis\textsuperscript{*}, Despoina Pavlidi\textsuperscript{*}}
\IEEEauthorblockA{\textsuperscript{*}Institute of Computer Science, Foundation for Research and Technology - Hellas, 70013 Heraklion, Crete, Greece\\
\textsuperscript{\dag}Department of Music Technology and Acoustics, Hellenic Mediterranean University, 74133 Rethymno, Greece\\
\small{\{nstefana, kalmar, andrsymi, stefpap, pavlidi\}@ics.forth.gr}
}}



\maketitle

\begin{abstract}
In this paper, we present an acoustic database, designed to drive and support research on voiced enabled technologies inside moving vehicles. The recording process involves (i) recordings of acoustic impulse responses, acquired under static conditions to provide the means for modeling the speech and car-audio components (ii) recordings of acoustic noise at a wide range of static and in-motion conditions. Data are recorded with two different microphone configurations,  particularly (i) a compact microphone array and (ii) a distributed microphone setup. We briefly describe the conditions under which the recordings were acquired, and we provide insight into a Python API that we designed to support the research and development of voice-enabled technologies inside moving vehicles. The first version of this Python API and part of the described dataset are available for free download.

\end{abstract}


\section{Introduction}
Over the last decade, interest in audio applications within vehicular environments has grown significantly. These applications range from speech recognition \cite{Hoshino_2004} to in-car audio perception \cite{paparas, Kaplanis_2017, Marchegiani_2022} and active noise control \cite{noiseControl}. A common prerequisite for all of these applications is the availability of relevant audio data, which has been addressed through various approaches in the literature.

Earlier works explored different methods for creating vehicular audio data sets. For example, in \cite{Saeta_2001}, speech components were recorded in an office and later combined with noise samples collected from a car environment. Similarly, in \cite{Mildner_2006} the authors used clean speech filtered with simulated impulse responses (IRs) and mixed with real car noise, while in \cite{Bisio_2018} clean speech was contaminated with simulated noise. Marques et al. \cite{Marques_2022} proposed a MEMS circular microphone array mounted at the car headliner for direction of arrival (DOA) estimation, using a controlled sine wave signal for measurements. Speech data in vehicles have also been captured using simpler setups, such as a single microphone mounted on the driver’s sun visor \cite{Hoshino_2004}.

The AVICAR audio-visual corpus \cite{Lee_2004} represents a more extensive dataset, containing 59,000 utterances from 100 different English speakers recorded with an 8-microphone linear array. This corpus is available upon request and remains one of the few comprehensive datasets in this domain. Other language-specific corpora, such as those for Czech \cite{Zelezny_2003}, Spanish \cite{Ortega_2004}, and Russian \cite{Kashevnik_2021}, have also been developed. Additionally, \cite{Soeta_2018} presented a sound field analysis by measuring head-related impulse responses (HRIRs) in vehicles, while \cite{Matheja_2013} used a four-microphone distributed array to record in-car speech and noise data.

 Most of these studies focus on specific scenarios, setups, or languages, with audio measurements typically limited to a single vehicle and relatively small datasets. Research has been mainly concentrated on driver-related audio content, neglecting the user experience from the perspective of the other passengers. Additionally, except from the AVICAR corpus \cite{Lee_2004}, most datasets are not widely available for further research or implementation \cite{Marchegiani_2022}.

In this paper, we present an acoustic database, along with a corresponding application programming interface (API), designed to advance research on voice-enabled technologies in vehicular environments. The purpose of the presented work and dataset is to serve as a fundamental resource, enabling rigorous experimentation and innovation in this field. In contrast to previous works, our dataset not only involves multiple different cars and microphone setups, but also recordings of real acoustic IRs from different passenger locations inside the cars. This approach offers unlimited flexibility, allowing users to synthesize audio scenarios with a customizable number of speakers and microphones, in any desired language and under various noise conditions. Users can incorporate their own clean (ideally anechoic) speech recordings, ensuring adaptability to diverse research and development scopes. 

The paper is structured as follows: The process to build the database is described in Section \ref{sec:main}, while in Section \ref{sec:synthesis}, we describe the basic approach behind the API which is designed to allow the user to easily produce a mixture containing components of speech and noise. Finally, in Section \ref{sec:metrics}, we present acoustic metrics associated with sound level and SNR as they were obtained under different driving conditions. Links to download a subset of the CAVEMOVE dataset and the corresponding API are provided at the end of this paper.

\section{DATA ACQUISITION PROCESS}
\label{sec:main}

\subsection{Microphone setups}
All recordings are acquired using M-audio M-Track Eight USB audio interface, at 48kHz sampling rate and at 24-bit sample size. Eight omnidirectional  microphones (Shure SM93) placed at different locations are used for recording acoustic noise and IRs. These are lightweight lavalier microphones with a flat frequency response from 80 to 20000 Hz. Two different microphone configurations have been used until now, as described below.\\
1) Microphone array: the eight microphones are mounted on a plastic circular case, forming a circular microphone array of radius equal to 5 cm. The circular array is placed on top of the dashboard, between the driver and the co driver (see Fig. \ref{fig:photos}(c)). Microphones were mounted on the array without a windshield.\\
2) Distributed  setup: The eight microphones are distributed throughout the car cabin, with six positioned at the front and two mounted on the headliner at the back (see Fig. \ref{fig:photos}(a) and (b)). While there are slight variations in placement between different cars, this setup ensures that at least one microphone is close to each passenger. All microphones were secured with blue tack and covered with a windshield.   

\subsection{Measurement of IRs}
\label{subsec:IRs}
Acoustic IRs describe the acoustic path characteristics between points of interest. In the construction of the CAVEMOVE database, IRs were recorded from each potential audio source (i.e., a speaker) to all microphones using the latest version of Room EQ Wizard (https://www.roomeqwizard.com/). A sweep tone served as the excitation signal. The measurements were conducted with the NTi TalkBox, a loudspeaker specifically designed to replicate the frequency response and radiation characteristics of the human voice. The loudspeaker was calibrated and featured built-in excitation signals, enabling the excitation of the car interior at acoustic power levels corresponding to specific speech efforts, namely, normal speech effort (referenced as 60 dBA at 1 m) and high speech effort (referenced as 70 dBA at 1 m) \cite{talkbox}. This process is of great importance as it enables the calculation of signal levels corresponding to specific speech efforts, thus facilitating the appropriate scaling of speech components. This ensures that, when combined with noise components, a realistic balance is achieved. The aforementioned process was repeated for each vehicle and microphone configuration, covering at least the driver, the front passenger, the rear-left, rear-middle, and rear-right passenger seats.

\begin{figure}[ht]
	\centering
	\includegraphics[width=0.94\linewidth]{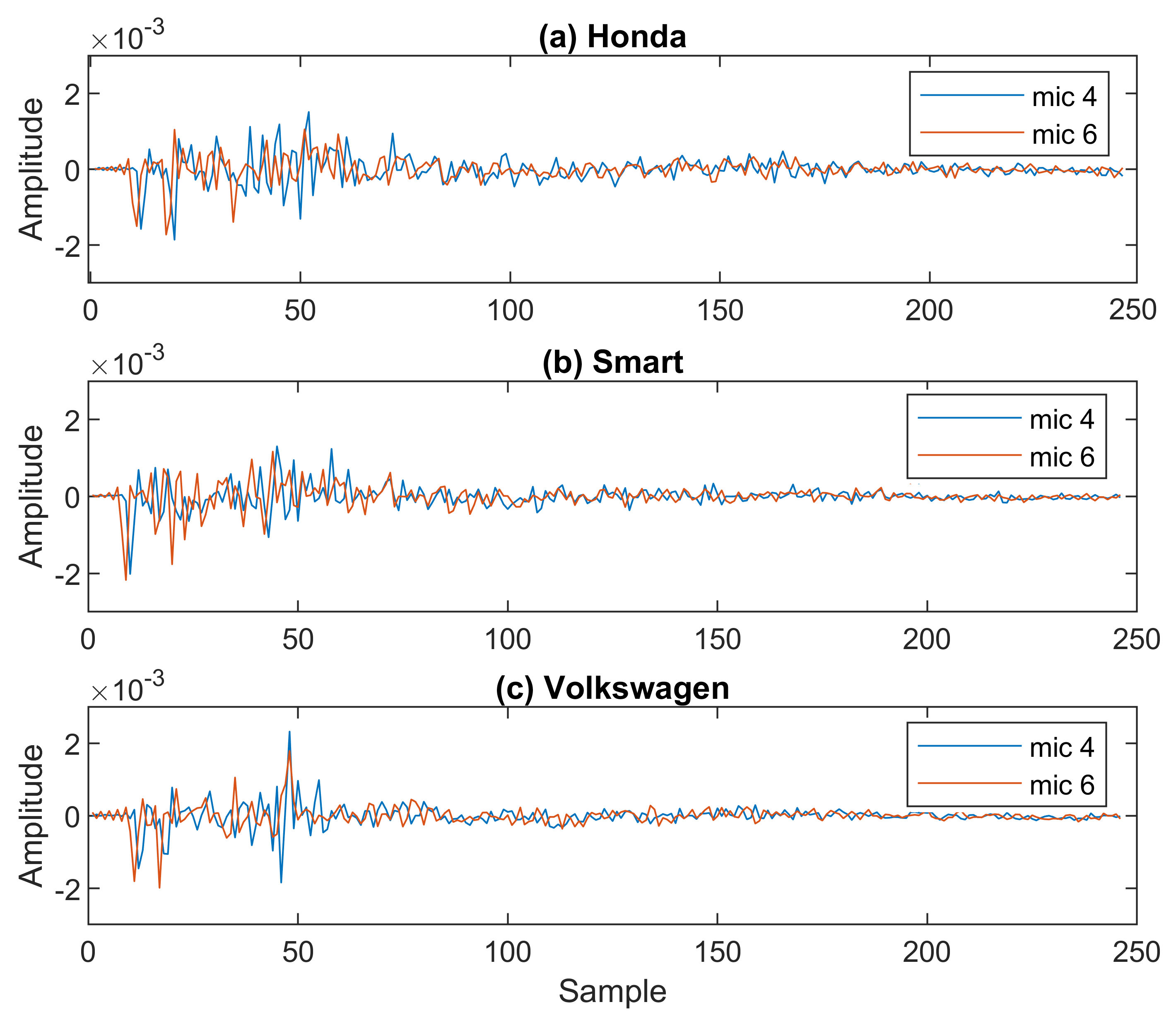}
	\caption{Plot of the IRs for microphones 4 and 6 when the car is excited from the driver's location. } 
\label{fig:IRs}
\end{figure}

In Fig. \ref{fig:IRs} we plot the IRs derived for microphone channels 4 and 6 when placed on the circular microphone array. The IRs are shown for (a) Honda, (b) Smart and (c) Volkswagen when the car is excited from the driver location. In all three cars the two front windows were slightly open (approximately 10 cm) and all other windows were closed. It can be observed that the direct path is directly followed by strong reflections which constitute a significant part of the energy in each IR. These reflections are comparable in strength to the direct path and seem to be related to the unique characteristics of the cabin of each car.

In addition to the acoustic paths from passenger locations to microphones, IRs are also measured for the built-in audio system, whenever available, recording the response from all car loudspeakers simultaneously. This was accomplished by directly feeding the excitation signal required for IR estimation into the auxiliary input of the audio system, or by reproducing it through the CD player. Attention was paid so that any adjustments related to equalization settings or panning were neutralized.

\subsection{Cars and acoustic conditions}
\label{subsec:conds}
So far, the recording process, as described in Section\,\ref{subsec:IRs}, has been repeated seven times, covering four different cars, as shown in Table\,\ref{tab:dataset}. Until now, more than fifteen hours of audio recording have been acquired and annotated, capturing a wide range of driving conditions and corresponding noise sources. 

In general, factors that affect the composition and level of acoustic noise at each microphone can be categorized into those that can be controlled and those that cannot be controlled. As controllable factors we consider car speed, window aperture and ventilation/air-condition level. On the other hand, non-controllable factors include weather condition, traffic, road surface (e.g., coarse, smooth, etc.) and external acoustic sources (e.g., vehicles, people, animals). Controllable factors have been attempted to be kept at fixed values for continuous temporal segments to achieve stationary noise conditions. In particular, regarding window aperture, we consider four states, as follows:\\
State 0: completely closed windows,\\ 
State 1: windows of the driver and front passenger being slightly open (approximately 10 cm),\\ 
State 2: completely open front windows and finally,\\
State 3: all four windows being completely open. 

Regarding the driving speeds, in each car we cover at least the range from 50 km/h to 110 km/h with steps of 10 km/h. All combinations of window apertures and driving speeds are covered, unless a combination is not feasible. In addition to these in-motion recordings, static recordings were obtained for capturing the noise produced by the built-in ventilation or air-conditioning system. Two or three different levels of ventilation power were considered in each car and the recording process was repeated so as to cover all combinations between the different ventilation levels and the different window apertures. 

Some additional conditions, captured with some but not all the cars and microphone setups, include e.g., speeds below 50 and above 110 km/h, driving on medium to highly coarse roads, and/or driving with rain or on a wet road. Inevitably, every recording session provides the opportunity to capture acoustic noise sources that are unpredictable or uncontrollable. These audio segments are annotated across a separate parallel layer so that they can be easily retrieved, with some of them being part of the basic dataset that is openly shared through this paper.

\begin{table}
	\begin{center}
		\begin{tabular}{lcccc}
			\hline
			\textbf{Brand}  & \textbf{Model} &  \textbf{Mic setup}   & \textbf{Audio system}  \\ 
			\hline
			Honda & CR-V (2009)& array \& distrib. & OK \\
			Alfa Romeo & 146 TS (2000) & distrib.  & - \\
			Smart &  forfour (2019) & array \& distrib. & OK \\
			Volkswagen & Golf (2011) & array \& distrib.  & OK  \\
			\hline
		\end{tabular}
	\end{center}
	\caption{Cars and microphone setups available until now in the CAVEMOVE dataset.}
	\label{tab:dataset}
\end{table}

\begin{figure*}[ht]
	\centering
	\includegraphics[width=0.98\linewidth]{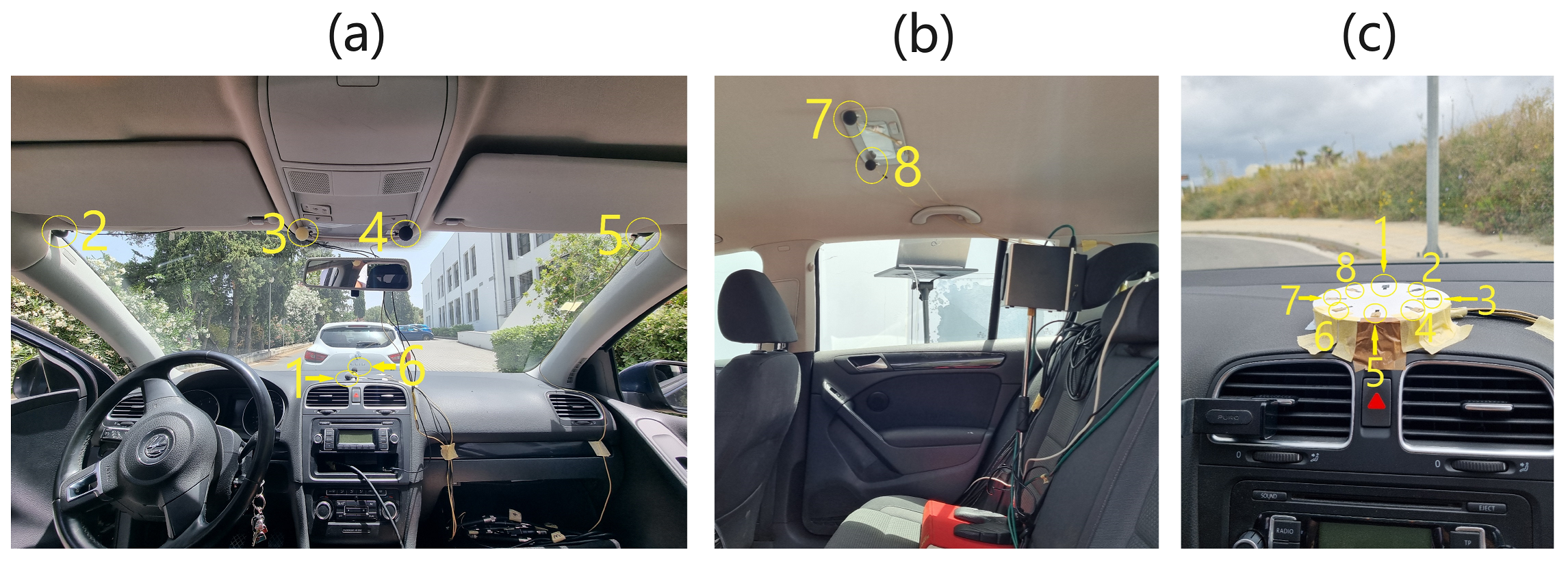}
	\caption{Photos with the microphone arrangement in VW. For the distributed setup, mics 1 to 6 at the front part of the car are shown in (a) and mics 7 and 8 at the back of the car in (b). The location of the microphone array is shown in (c). The loudspeaker used for obtaining the IRs can also be seen in (b), at the rear-right passenger seat.} 
\label{fig:photos}
\end{figure*}
\subsection{Additional notes}
\label{subsec:additional}

In general, the data acquisition process that was repeated with each car and microphone setup lasted more than one day. During this period, the microphones were constantly fixed to their initial locations and attention was paid so as to avoid any accidental displacements. The potentiometers of the sound card, which affect the pre-amplification gain associated to each microphone channel, were also fixed. Moreover, before and after every recording session, each microphone channel was subjected to an empirical calibration process which would allow us to derive an approximate estimation of the free field sensitivity for each input channel. Calibration took place inside the FORTH-ICS semi-anechoic chamber; each microphone, while still connected to its nominal soundcard input, was placed at a specific point inside the chamber and was subjected to pink noise, emitted from one of the loudspeakers present in the lab. Subsequently, a Class-II omnidirectional microphone (intended for acoustic measurements) that was connected to an NTi sound level meter was placed at the same location and subjected to the same pink noise emission. The A-weighted Equivalent Noise Level of the measurement microphone was noted down. Subsequently, each recorded pink noise signal was passed through an IIR filter that exhibits the characteristic frequency response of the A-weighted filter \cite{Beranek}. Through this process, we were able to associate recorded signal levels to sound pressure level in dB$A$, and thus derive an approximate estimation of each microphone's sensitivity. We note that while the estimated sensitivity is only approximately correct, the accuracy achieved is considered sufficient for the needs of this work.

\section{SYNTHESIS MODEL}
\label{sec:synthesis}

The ultimate goal of the data and python API delivered in the context of CAVEMOVE is to allow the engineers and researchers to easily synthesize the microphone signals that correspond to a particular scenario. Assuming that a target car and microphone setup has been chosen, the synthesis process can be compactly described as
\begin{equation}
	\mathbf{Y}=\mathbf{S}(p,L_s,w,\mathbf{x})+\mathbf{A}(L_a,w,\mathbf{z})+\mathbf{N}(s,w)+\mathbf{V}(l,w).
	\label{eq:synthesisModel}
\end{equation}
Briefly, each bold capital symbol is a $N \times M$ signal matrix, where $M$ is the desired number of microphone channels and $N$ is the duration of the synthesized sound excerpt in samples. $\mathbf{S}$ represents the filtered speech components, $\mathbf{A}$ represents the interference components produced by the built-in audio system (when available),  $\mathbf{N}$ are the noise components corresponding to the particular driving condition and $\mathbf{V}$ are the noise components associated to the ventilation/air-condition functionality. Assuming that all these sound components are independent from one another, the final microphone signal can be synthesized as in Eq. (\ref{eq:synthesisModel}), by means of simple superposition. As it can be seen, each component is produced as a function of user defined parameters. A brief explanation of these parameters is as follows; 

\begin{table*}[ht!]
	\centering
	\begin{tabular}{|c|c|c|c|c|c|c|c|c|c|c|}
		\hline
		\multicolumn{3}{|c|}{Condition}  & \multicolumn{2}{|c|}{Alfa Romeo} & \multicolumn{2}{|c|}{VW} & \multicolumn{2}{|c|}{Honda}  & \multicolumn{2}{|c|}{Smart}\\ \hline
		speed  & window & ventilation & SNR & Noise level & SNR  & Noise level & SNR  & Noise level & SNR  & Noise level\\ 
		(km/h) & (w) & & (dB) & (dBA) & (dB) & (dBA) & (dB) & (dBA) & (dB) & (dBA)\\ \hline
		0 & 0 & off & 17.9 & 49.2 & 21.1 & 42.6 & 21.8 & 42.7 &16.2 & 46.9\\ \hline
		0 & 2 & off & 13.7 & 52.9 & 18.5 & 45.4 & 18.0 & 45.8 & 15.1 & 48.0\\ \hline
		0 & 0 & level = 1 & 11.6 & 55.5 & 11.7 & 52.7 & 15.3 & 49.2 & -1.5 & 64.6\\ \hline
		0 & 0 & level = 2 & -0.2 & 67.3 & 0.2 & 64.2 & 3.8 & 60.7 & -8.9 & 72.0\\ \hline
		0 & 0 & level =  3 & N/A & N/A & -7.1 & 71.5 & -2.5 & 67.0 & -14.3 & 77.4\\ \hline
		40 & 0 & off & 1.5 & 65.6 & 4.3 & 60.1 & 3.3 & 61.2 & 1.6 & 61.5\\ \hline
		50 & 2 & off & -5.1 & 71.7 & -3.1 & 67.0 & -7.5 & 71.3 & -4.2 & 67.4\\ \hline
		70 & 0 & off & -3.4 & 70.5 & 1.0 & 63.4 & -2.0 & 66.5 &-2.1 & 65.2\\ \hline
		70 & 1 & off & -5.6 & 72.7 & -6.8 & 71.0 & -6.0 & 70.0 & -5.4 & 68.4\\ \hline
		70 & 3 & off & N/A & N/A & -7.0 & 71.0 & -10.5 & 74.4 & N/A & N/A\\ \hline
		80 & 2 & off & -11.1 & 77.8 & -7.5 & 71.4 & -12.2 & 76.0 & -8.6 & 71.7\\ \hline
		90 & 0 & off & -3.4 & 70.5 & -1.1 & 65.5 & -4.0 & 68.5 & -3.4 & 66.5\\ \hline
		100 & 1 & off & -8.8 & 75.8 & -8.8 & 72.9 & -10.1 & 74.1 & -10.4 & 73.4\\ \hline
		100 & 2 & off & -13.3 & 79.9 & -11.3 & 75.2 & -14.0 & 77.8 & -11.9 & 75.0\\ \hline
		110 & 2 & off & -17.1 & 83.8 & -13.4 & 77.3 & -14.5 & 78.3 & -15.5 & 78.6\\ \hline
		110 & 3 & off & N/A & N/A & -17.0 & 81.1 & -13.6 & 77.5 & N/A & N/A\\ \hline
	\end{tabular}
	\caption{Noise level and SNR for different driving conditions inside the four different vehicles. All values are measured on a reference microphone. SNR is calculated for normal speech effort with the speech sound source located at the driver's seat.}
	\label{tab:dBs}
\end{table*}

\begin{itemize}
	\item $w$: is an integer taking values in the range $[0,1,2,3]$, with each integer value representing a different condition with respect to the windows' apertures, as described in Section \ref{subsec:conds}.
	\item $\mathbf{x}$: is a one-dimensional vector with the dry (ideally anechoic) speech recording that is used for the particular scenario. The user is responsible for providing an appropriate speech recording in PCM format. The audio recording can be of any length and sampling rate (it will be automatically converted to the target sampling rate).
	\item $p$: it can be selected from a given set of string values, each value corresponding to a particular location of the passenger inside the car cabin. For each combination of $p$ and $w$ the corresponding impulse response is automatically loaded and used to produce the filtered speech components by means of convolution with $\mathbf{x}$.
	\item $L_s$: it is a user defined value, in dBA, corresponding to the speech effort. While this can be any non-negative value, we recommend values in the range between 60 and 70, with 60 corresponding to normal speech effort and 70 to a high speech effort \cite{talkbox}.
	\item $L_a$: is a user defined value, in dBA, corresponding to the mean A-weighted acoustic level of the audio program reproduced from the built-in audio system. The corresponding interference components are automatically scaled as a function of this value so that the signal level at a reference microphone matches the desired sound level. Again, any non-negative value is applicable and it is up to the user to set it to a realistic value.
	\item $\mathbf{z}$: is a one-dimensional vector representing the audio program. The user is responsible for providing an appropriate audio file in PCM format. The audio file can be of any length and sampling rate. Audio files with more than one channels are accepted but will be automatically downmixed to a monophonic audio signal.
	\item $s$: is an integer value within a predefined set of different speeds in km/h. The noise components corresponding to particular speed, window aperture, car and microphone setup are automatically loaded.
	\item $l$: it is an integer with values in the range $[1,2,3]$, each value corresponding to a different level of the ventilation/air-conditioning system, so that higher levels produce more noise.
\end{itemize}     

The synthesis approach is designed in such way that the resulting length $N$ of the output signal $\mathbf{Y}$, as well as of all matrices in Eq. (\ref{eq:synthesisModel}) matches the length of the dry speech signal $\mathbf{x}$ when converted to the target sampling rate. To achieve this, we have incorporated a mechanism that recycles the noise sequences inherent to the construction of $\mathbf{A}$, $\mathbf{N}$ and $\mathbf{V}$ as many times necessary in order to match the length of $\mathbf{x}$. Note that in most cases this will not be necessary if the dry speech signal does not exceed 25 s of duration.  

The presented synthesis approach provides an easy way for synthesizing mixtures which preserve a realistic balance between speech and noise components without the need for the end-user to manually scale the signal matrices. While our approach does not allow to explicitly define the SNR, by changing the values of $L_s$, $L_a$, $w$, $s$ and $l$ the user can simulate a wide range of conditions in terms of the balance between the speech and noise components. We believe that this approach allows for more realistic simulation, by preserving the spectral profiles of the noise sources corresponding to specific driving conditions and SNRs. Subsequently, performance metrics derived using CAVEMOVE dataset and API are expected to better reflect the actual performance when the studied methods are deployed in real conditions. 

Additional to the above functionalities, the API includes auxiliary functions, as for example a function that returns steering vectors applicable to the circular microphone array and a function that returns the impulse responses involved with a particular scenario. These functionalities might help the user towards the design of acoustic beamformers and post filters required for deploying typical speech enhancement algorithms. The audio data that we openly share with this paper is in PCM format, at 16 kHz sampling rate and at 24 bits. Interested readers can download the python API from CAVEMOVE's GitHub account \cite{github}, where they will also find a link to download the audio data.

\section{Acoustic metrics}
\label{sec:metrics}

We provide average noise sound levels and SNRs for different driving conditions in the four vehicles in Table \ref{tab:dBs}. These values are calculated on a specific microphone in each car, at the original sampling rate of 48 kHz. Particularly, in Volkswagen and Smart the values are obtained using microphone 5, placed as depicted in Fig. \ref{fig:photos}(c), while in Honda and Alfa Romeo, the microphone taken into account is placed approximately as microphone 3 in Fig. \ref{fig:photos}(a). The average noise sound levels are obtained using the A-weighted free field sensitivity value of the reference microphone, as it was calculated following the procedure described in Section \ref{subsec:additional}. On the other hand, SNR is calculated by subtracting the A-weighted noise level by the A-weighted speech sound level corresponding to normal vocal effort, as it was calculated by placing the calibrated loudspeaker on the driver's seat. 

As expected, the noise level increases with car speed and window aperture. Starting from 42.0 dBA at 0 km/h, the noise level exceeds 80 dBA at the highest speeds in certain occasions. This is of course also reflected on the SNR, which spans a range that in all cases exceeds 30 dB between the quietest and loudest condition. It is interesting to observe that, even in moderate conditions, e.g. at 70 km/h with closed windows, the SNR can drop to values below 0 dB. We note again that the presented metrics are derived for normal speech effort and we do not account for the fact that instinctively, the speaker will likely raise the level of his/her voice in loud noise conditions \cite{JASA01}. It is also interesting to note that, even in idle conditions (0 km/h) and closed windows (w=0), the noise produced by the car ventilation system can easily reduce SNR to values close to or below 0 dB. 

It is tempting to try to spot differences among the different cars. For example, Smart has by far the noisiest ventilation system among all cars. Also, it appears that Alfa Romeo, being the oldest among all cars, is also the noisiest one. However, to be fair, it should be mentioned that noise levels corresponding to conditions where the windows are open are highly affected by the environmental conditions and especially by the speed of wind and its direction with respect to the direction of the car. 

\section{Conclusion and future work}
The study of voice-enabled technologies inside moving vehicles is an important prerequisite for improving the user experience in the cars of the future. CAVEMOVE is a research project dedicated to the collection of audio data to facilitate this study. Until now, data have been collected inside four vehicles, involving two different microphone setups and accounting for acoustic noise sources associated not only to the car's motion, but also to the built-in audio system and the ventilation/air-conditioning system. A carefully designed API is provided for free download, alongside with the necessary audio data, for the user to easily synthesize the sound signals corresponding to a wide range of different driving conditions. We believe that this dataset can be a very valuable tool for evaluating performance of different algorithms but also for training machine learning models for applications such as voice activity detection, speech enhancement, speech separation, speech/speaker recognition and speaker localization. In the future, we plan to extend this dataset with additional recordings and cars and furthermore, we plan to make available a Matlab version of the API.  


\end{document}